\documentclass[aps,prl,floatfix,twocolumn,reprint,amsmath,amssymb,superscriptaddress,UTF8]{revtex4-2}
\usepackage{amsfonts}
\usepackage{mathrsfs}
\usepackage{amsmath}
\usepackage{color}
\usepackage{natbib}
\usepackage{graphicx}
\usepackage{bm}
\usepackage{amssymb}
\usepackage{xspace}
\usepackage{epstopdf}
\usepackage{dcolumn}
\usepackage{longtable}
\usepackage{array}
\usepackage{makecell}
\usepackage{tabulary}
\usepackage{multirow}
\newcolumntype{C}[1]{>{\centering\arraybackslash}p{#1}}
\newcolumntype{L}[1]{>{\flushleft\arraybackslash}p{#1}}

\usepackage{multirow}
\usepackage{epsfig}
\usepackage[normalem]{ulem}
\usepackage{amsmath}
\usepackage{braket}
\usepackage{bbold}
\usepackage{natbib}

\usepackage[colorlinks=true, letterpaper=true, pdfstartview=FitV, linkcolor=blue, citecolor=blue, urlcolor=blue]{hyperref}

\makeatletter

\newcommand{\Rmnum}[1]{\expandafter\@slowromancap\romannumeral #1@}
\makeatother

\begin{document}

\title{Quantized Spin-Hall Conductivity in Altermagnet Fe$_2$Te$_2$O with Mirror-Spin Coupling}

\author{Run-Wu Zhang}
\thanks{These authors contributed equally to this work.}
\affiliation{Key Lab of advanced optoelectronic quantum architecture and measurement (MOE), Beijing Key Lab of Nanophotonics $\&$ Ultrafine Optoelectronic Systems, and School of Physics, Beijing Institute of Technology, Beijing 100081, China}
\affiliation{International Center for Quantum Materials, Beijing Institute of Technology, Zhuhai, 519000, China}

\author{Chaoxi Cui}
\thanks{These authors contributed equally to this work.}
\affiliation{Key Lab of advanced optoelectronic quantum architecture and measurement (MOE), Beijing Key Lab of Nanophotonics $\&$ Ultrafine Optoelectronic Systems, and School of Physics, Beijing Institute of Technology, Beijing 100081, China}
\affiliation{International Center for Quantum Materials, Beijing Institute of Technology, Zhuhai, 519000, China}

\author{Yang Wang}
\affiliation{Key Lab of advanced optoelectronic quantum architecture and measurement (MOE), Beijing Key Lab of Nanophotonics $\&$ Ultrafine Optoelectronic Systems, and School of Physics, Beijing Institute of Technology, Beijing 100081, China}
\affiliation{International Center for Quantum Materials, Beijing Institute of Technology, Zhuhai, 519000, China}

\author{Jingyi Duan}
\affiliation{Key Lab of advanced optoelectronic quantum architecture and measurement (MOE), Beijing Key Lab of Nanophotonics $\&$ Ultrafine Optoelectronic Systems, and School of Physics, Beijing Institute of Technology, Beijing 100081, China}
\affiliation{International Center for Quantum Materials, Beijing Institute of Technology, Zhuhai, 519000, China}

\author{Zhi-Ming Yu}
\email{zhiming\_yu@bit.edu.cn}
\affiliation{Key Lab of advanced optoelectronic quantum architecture and measurement (MOE), Beijing Key Lab of Nanophotonics $\&$ Ultrafine Optoelectronic Systems, and School of Physics, Beijing Institute of Technology, Beijing 100081, China}
\affiliation{International Center for Quantum Materials, Beijing Institute of Technology, Zhuhai, 519000, China}

\author{Yugui Yao}
\email{ygyao@bit.edu.cn}
\affiliation{Key Lab of advanced optoelectronic quantum architecture and measurement (MOE), Beijing Key Lab of Nanophotonics $\&$ Ultrafine Optoelectronic Systems, and School of Physics, Beijing Institute of Technology, Beijing 100081, China}
\affiliation{International Center for Quantum Materials, Beijing Institute of Technology, Zhuhai, 519000, China}
\date{\today}
\begin{abstract}
Due to  spin-orbit coupling (SOC), crucial for the quantum spin Hall (QSH) effect, a quantized spin-Hall conductivity has not yet been reported in QSH insulators and other  realistic  materials.
Here, we tackle this challenge by predicting robust quantized  spin-Hall conductivity in monolayer Fe$_2$Te$_2$O.
The underlying physics originates from the  unrecognized mirror-spin coupling  (MSC), which couples spin-up and spin-down states into  two orthogonal mirror eigenstates.  
We show  that the MSC can naturally emerge in the two-dimensional altermagnets with  out-of-plane N\'eel vector and  horizontal mirror.
A remarkable consequence of the MSC is that it can dramatically weaken the spin hybridization of the altermagnetic materials when  SOC is included.
When SOC is neglected, Fe$_2$Te$_2$O is an altermagnetic Weyl semimetal  with MSC.
With SOC,  it evolves into  the first material candidate for    magnetic mirror Chern insulator.
Remarkably, under the protection of MSC, the spin hybridization of  both bulk and topological edge states in Fe$_2$Te$_2$O with SOC at low energy is negligible.
As a consequence, a quantized  spin-Hall conductivity emerges within the bulk band gap of the system.
By unveiling a novel  effect, our findings  represent a significant advancement  in spin Hall transport, and broaden the material candidates  hosting  intriguing altermagnetic phenomena.
\end{abstract}
\maketitle

\textit{\textcolor{blue}{Introduction.}}--
Topological materials have garnered significant interest and rapidly advanced in both fundamental research and practical applications~\cite{novoselov2005two, mak2010atomically, li2014black, liu2011quantum, zhao2016rise, xu2013large, reis2017bismuthene, gong2017discovery, chang2013experimental, li2017evidence, deng2020quantum, li2019intrinsic}. 
These systems are mathematically characterized by quantized topological charge, such as the Chern number in quantum anomalous Hall (QAH) insulators  and the $Z_2$ invariant in quantum spin Hall (QSH) insulators.
Many efforts have been made to unveil quantized physical quantities associated with topological charge~\cite{de2017quantized, liu2020quantized, li2025planar}.
A typical example is the Hall conductivity of QAH insulators, which is directly proportional to the Chern number of the systems and thus becomes quantized~\cite{bernevig2013topological}. 

Similarly, one may expect that a quantized spin-Hall conductivity could be achieved in QSH insulators.
However, this is generally  impossible due to a fundamental restriction~\cite{maciejko2011quantum}.
To realize a significant  QSH effect, the system should exhibit strong spin-orbit coupling (SOC)~\cite{hasan2010colloquium, qi2011topological}.
However, the SOC will  hybrid the spin-up and spin-down states,  making each of the topological helical edge states in the QSH insulator lack persistent spin polarization.
Therefore, for a  four-terminal  setup, which   is  used to  measure the spin-Hall transport, the spin-Hall conductivity of the QSH insulator generally is not a  constant within the bulk band gap~\cite{maciejko2011quantum}.
Consequently, while quantized spin-Hall conductivity can be found  in many simplified   effective models~\cite{kane2005z, bernevig2006quantum, zou2024topological, antonenko2024mirror}, where the spin is a conserved quantity, it remains a challenge to identify material candidates that host this quantized effect, as SOC is an intrinsic property of realistic materials.

\begin{figure}
\includegraphics[width=1.0\columnwidth]{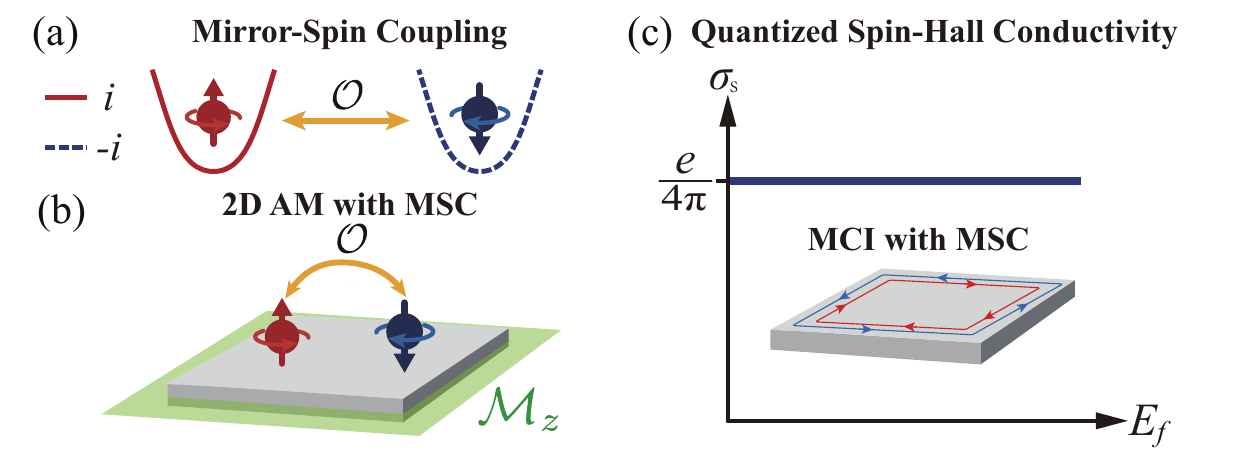}
\caption
{(a) Illustration of MSC. With MSC, the two symmetry-connected bands with opposite spin polarizations are endowed with opposite mirror eigenvalues. (b) 2D AM systems with  out-of-plane N\'eel vector and horizontal mirror naturally have MSC. 
(c) An MCI with MSC hosts counter-propagating topological edge states comprising electrons of opposite but persistent spin polarization, leading to quantized spin-Hall conductivity. In (a), solid and dashed lines respectively show the opposite mirror eigenvalues.}
\label{fig1}
\end{figure}

Here, we address this challenge by going beyond the traditional QSH effect, circumventing this fundamental restriction with the help of an unrecognized coupling effect.
The  physics of this work is to explore  a two-dimensional (2D)  mirror Chern insulator (MCI) with mirror-spin coupling (MSC).
For a 2D system with a horizontal mirror plane (${\cal{M}}_z$), each energy  band is endowed with a certain mirror eigenvalue.
The MSC  effect here  means that the spin polarization of the states in each band depends on the  mirror eigenvalue of the band, and the two symmetry-connected bands with different mirror eigenvalues exhibit opposite spin polarizations, as shown in  Fig. \ref{fig1}(a). 
Via symmetry analysis, we first show that the MSC naturally appears in the  recently proposed  altermagnetic (AM)  systems~\cite{vsmejkal2020crystal, gonzalez2021efficient, smejkal2022beyond, smejkal2022emerging, smejkal2022giant, ma2021multifunctional, shao2023neel, bai2024altermagnetism} with out-of-plane N\'eel vector and horizontal mirror [see Fig. \ref{fig1}(b)].
In AM materials, the spin is a conserved quantity, but is split in momentum space by the Zeeman effect, leading to various unique properties~\cite{Bai2022observation, lee2024broken, guo2023quantum, guo2023altermagnetic, guo2025luttinger, zhang2024predictable, he2024quasi, liu2024twisted, gu2024ferroelectric, sun2024stacking}.
A major advantage of altermagnets is the decoupling of spin-up and spin-down states.
However, this advantage generally is   diminished by the SOC effect.
Surprisingly, we find that the MSC can significantly reduce the SOC-induced spin hybridization, allowing the SOC systems still to have persistent spin polarization.
Therefore, for  an AM material with MSC, when it evolves into an MCI under the SOC effect, we will obtain an MCI with MSC (see  Fig. \ref{fig1}).
Then, one can expect that  the two counter-propagating topological edge states of this novel MCI should exhibit  opposite but persistent spin polarization, leading to quantized spin-Hall conductivity in the bulk band gap, as illustrated in Fig. \ref{fig1}(c).

Based on first-principle calculations, we demonstrate our ideas by  predicting  the monolayer Fe$_2$Te$_2$O.
Without SOC, the  Fe$_2$Te$_2$O is an AM material with MSC, and has two ideal Weyl points (WPs) in each spin channel around the Fermi level.
Due to  MSC, the spin-opposite WPs have different  ${\cal{M}}_z$ eigenvalues $m_z=\pm i$.
Particularly, they are well separated from each other in $k$-space, and also are isolated  from other bands in energy.
When SOC is included, all the WPs are gapped out. 
The gapped  WPs with $m_z=i$ and $m_z=-i$ respectively give a Chern number of $1$ and $-1$, making monolayer Fe$_2$Te$_2$O  the  first material candidate of  magnetic MCI.  
Notably, the spin hybridization of  both bulk and topological edge states in Fe$_2$Te$_2$O with SOC at low energy is negligible, as protected by MSC.
As a consequence, a quantized spin-Hall conductivity indeed is observed in the bulk band gap of the system.
Our work  advances our fundamental understanding of topological phases in AM systems, and paves the way for the development of quantized spintronics  devices.

\textit{\textcolor{blue}{MSC in Altermagnets.}}--
We begin by demonstrating that  the MSC effect can naturally arise in the 2D AM materials with  out-of-plane N\'eer vector and  horizontal mirror, as shown in Fig. \ref{fig1}(b).

The symmetry of AM systems is described by the  spin space group (SSG)~\cite{brinkman1966theory, litvin1974spin, brinkman1966space, xiao2024spin, chen2024enumeration, jiang2024enumeration}.
Since the spin direction and spatial coordinates in SSG  vary independently, the ${\cal{M}}_z$ in this AM system can be  written as ${\cal{M}}_z=\{M_z^s||M_z^l\}=\{C_{2z}^sP^s||M_z^l\}$, where $C_{2z}^{s(l)}$, $P^{s(l)}$ and $M_z^{s(l)}$ represent the two-fold rotation, inversion and horizontal mirror acting on spin (lattice) space, respectively.  Notice that in spin space, $P^s$ is identical to the identity operator $E^s$.
Therefore, we can express the mirror symmetry as ${\cal{M}}_z=\{C_{2z}^s||M_z^l\}$, with eigenvalues being $m_z=\pm i$ due to the presence of spin operator.
Besides, the system must have (at least) one  operator, denoted as  ${\cal{O}}$, that connects the two different  spins.
The ${\cal{O}}$ can be either ${\cal{O}}=\{C_{2,\|}^s||O^l\}$ where $C_{2,\|}^s$ is an arbitrary two-fold in-plane rotation on spin, or $\{E||O^l\}{\cal{T}}=\{C_{2,\|}^s||O^l\}*\{C_{2,\|}^s|E^l\}{\cal{T}}$ with  $\{C_{2,\|}^s|E^l\}{\cal{T}}$  being an intrinsic symmetry of  AM. Here, ${\cal{T}}$ is the time-reversal symmetry.

For simplicity, we use point spin group symmetry to discuss the MSC effect.
Since  $C_{2,\|}^s$ anticommutes with $C_{2z}^{s}$, and all symmorphic operators $O^l$ commute with $M_z^l$, we  always have 
\begin{eqnarray}\label{AM}
\{{\cal{M}}_z, \ {\cal{O}}\} & =& 0,
\end{eqnarray}
for ${\cal{O}}=\{C_{2,\|}^s||O^l\}$, and 
\begin{eqnarray}\label{AM2}
[{\cal{M}}_z, \ {\cal{O}}] & =& 0,
\end{eqnarray}
for ${\cal{O}}=\{E||O^l\}{\cal{T}}$.

The Bloch states of the AM system in  the Brillouin zone (BZ) can be chosen as the eigenstates of  ${\cal{M}}_z$, which we denoted as $|m_z,{\bm k}_{\uparrow(\downarrow)}\rangle$ with $m_z=\pm i$, and  ${\bm k}_{\uparrow(\downarrow)}$ representing the momentum of the  spin-up (spin-down) state.
From Eq.~(\ref{AM}), one finds that 
\begin{eqnarray}\label{Eq2}
{\cal{M}}_z{\cal{O}}|\pm i,{\bm k}_{\uparrow(\downarrow)}\rangle & =& \mp i {\cal{O}}|\pm i,{\bm k}_{\uparrow(\downarrow)}\rangle, 
\end{eqnarray}
for both ${\cal{O}}=\{C_{2,\|}^s||O^l\}$ and ${\cal{O}}=\{E||O^l\}{\cal{T}}$.
This indicates that ${\cal{O}}|\pm i,{\bm k}_{\uparrow(\downarrow)}\rangle=|\mp i,{\bm k}^{\prime}_{\downarrow(\uparrow)}\rangle$, where ${\bm k}'=O^l {\bm k}$ or  ${\bm k}'=-O^l {\bm k}$ (when ${\cal{O}}$ includes ${\cal{T}}$).
Consequently, the symmetry-connected spin-up and spin-down bands should have opposite mirror eigenvalues, resulting in the MSC in AM systems.

\begin{figure}
\includegraphics[width=1.0\columnwidth]{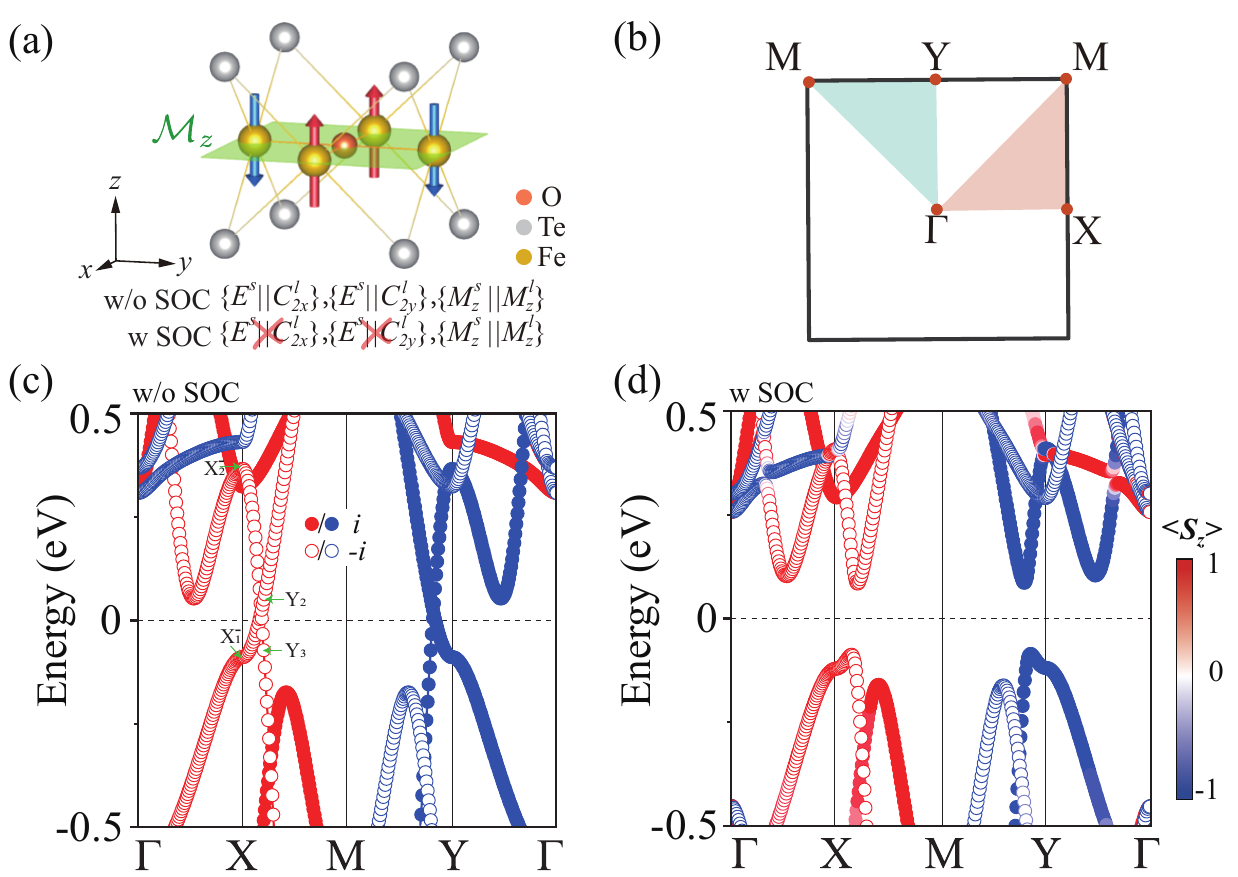}
\caption
{(a) shows the relaxed geometry of monolayer Fe$_2$Te$_2$O, which exhibits  out-of-plane N\'eel vector  and  ${\cal{M}}_z$. The ${\cal{M}}_z$ is always a symmetry operator of Fe$_2$Te$_2$O, regardless of the SOC. (b) Brillouin zone (BZ) of   Fe$_2$Te$_2$O.
(c-d) Electronic band structure of  Fe$_2$Te$_2$O (c) without (w/o) and (d) with (w) SOC. 
In (c), the bands in spin-up and spin-down  channels are plotted as  red and blue curves, respectively. 
In (d), the colors indicate out-of-plane spin polarization ($s_z$).
In (c-d), solid and hollow circles respectively denote the states with ${\cal{M}}_z=+i$ and ${\cal{M}}_z=-i$.}
\label{fig2}
\end{figure}

The MSC can lead to an important consequence for the AM systems. 
Most of the unique phenomena in AM systems arise from the decoupling of the spin-up and spin-down states, a feature that can be diminished by SOC. 
However, the MSC provides a protection for this decoupling, as ${\cal{M}}_z=\{M_z^s||M_z^l\}$ persists even under the influence of SOC.
This protection is  most significant when  the electronic bands in each spin channel near the Fermi level share the same mirror eigenvalues, while all other bands lie far away from the Fermi level [see Fig. \ref{fig2}(c)].
In such a case, the hybridization  between the spin-up and spin-down states at low energies is negligible.
As a result, systems with MSC can maintain persistent spin polarization despite the presence of SOC. 
This means that the MSC systems can be  ideal platforms for combining two conflicting effects:  SOC and spin decoupling (persistent spin texture), which is a critical challenge in the field of spintronics~\cite{schliemann2017colloquium, tao2018persistent, zhao2020purely, garcia2020canted, krol2021realizing, tian2017observation}, and can  effectively  address the aforementioned fundamental restriction for achieving quantized spin Hall conductivity (provided  the system is a magnetic MCI).

We note that Eqs.~(\ref{AM}-\ref{Eq2}) also may be satisfied in certain SOC systems. However, this can not guarantee these SOC systems to have  MSC, due to the spin hybridization.  To achieve MSC in SOC systems, the practical way is finding  AM materials satisfying Eqs.~(\ref{AM}-\ref{Eq2}) and maintaining a clean band structure around the Fermi level. 
 
\textit{\textcolor{blue}{MSC  material candidates.}}--
In physics, many promising concepts have made significant theoretical progress. However, turning these ideas into practical material systems  depends heavily on the discovery of suitable candidate materials.
Here, we demonstrate that monolayer Fe$_2$Te$_2$O is a candidate with MSC.
 
Monolayer Fe$_2$Te$_2$O can be  obtained through mechanical exfoliation from bulk  Fe$_2$Te$_2$O~\cite{refsm}, which is a homologue of experimentally  synthesized materials: layered iron oxytelluride family BaFe$_2$Ch$_2$O (where Ch represents a chalcogen)~\cite{takeiri2016high, song2023crystal, he2011synthesis}. 
The crystal structure of the monolayer Fe$_2$Te$_2$O consists of three atom layers arranged in the sequence Te-(Fe-O)-Te, as shown in Fig. \ref{fig2}(a). The  two Fe atoms and one O atoms lie in the same plane, sandwiched between the two Te atomic layers. The crystal lattice belongs to  the space group P4/mmm (No. 123), with an optimized lattice constant of $a=b=4.03$ \AA.
Phonon spectrum and molecular dynamics simulations further confirm the  structural stability \cite{refsm}, indicating that Fe$_2$Te$_2$O is robust enough to form a freestanding 2D material.
By comparing the energies of several possible magnetic configurations, we confirm that the ground state of the monolayer Fe$_2$Te$_2$O exhibits an AM configuration~\cite{refsm}, as illustrated in Fig. \ref{fig2}(a). The magnetic moments localized on the Fe sites, with a magnitude of $\sim$4 $\mu_B$, and align along the out-of-plane easy axis, resulting in a maximum value of the magnetic anisotropy energy of 7.014 meV/Fe~\cite{refsm}. 
Without  SOC, monolayer Fe$_2$Te$_2$O belongs to the SSG No. $47.123.1.1$, having the horizontal mirror ${\cal{M}}_z=\{M_z^s||M_z^l\}$.
The sublattices with opposite spins are connected by $\{E^s||C_{4z}^l\} \mathcal{T}$ and ${\cal{M}}_{\bar{1}10/110}=\{C_{2,\bar{1}10/110}^s||M_{\bar{1}10/110}^l\}$.
From Eqs. (\ref{AM2}) and (\ref{Eq2}), one knows that  $\{E^s||C_{4z}^l\} \mathcal{T}$ would invert the eigenvalue of ${\cal{M}}_z$.
Therefore, the electronic states connected by $\{E^s||C_{4z}^l\} \mathcal{T}$ will  have opposite both spin and ${\cal{M}}_z$ eigenvalues, leading to the MSC.

\begin{figure}
\includegraphics[width=1.0\columnwidth]{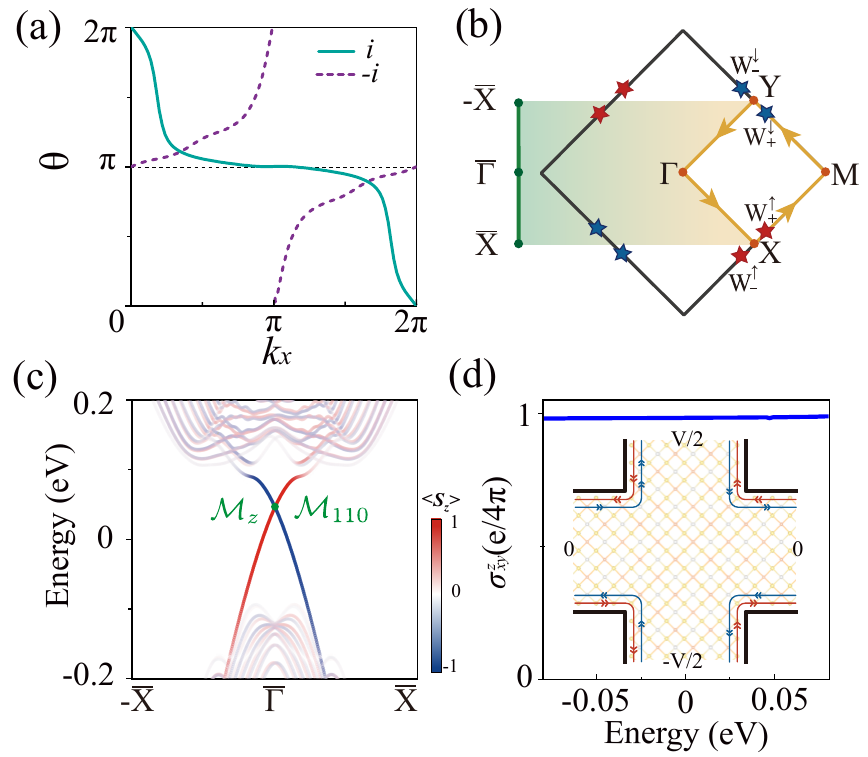}
\caption
{(a) Wilson loops of the  Fe$_2$Te$_2$O with SOC, which are obtained in the two ${\cal{M}}_z$ (${\cal{M}}_z=\pm i$) subsystems, respectively. (b) Bulk and edge BZs. The location of the four AM WPs in Fe$_2$Te$_2$O without SOC are presented in bulk BZ.
(c) The spectrum of the nanoribbon with $(1\bar{1}0)$ edges. The  colors denote the  out-of-plane spin polarization. The topological  edge states exhibit an almost 100\% spin polarization. 
The band crossing of the edge states at ${\bar \Gamma}$ point is doubly protected by ${\cal{M}}_z$ and ${\cal{M}}_{110}$.
(d) The spin-Hall conductivity $\sigma_{xy}^z$ in the bulk band gap, obtained by a four-terminal measurement setup. 
Due to  the almost perfect spin polarization of the  edge states, the spin-Hall conductivity $\sigma_{xy}^z$ is quantized to the expected  constant $e/(4\pi)$.
}
\label{fig3}
\end{figure}

\textit{\textcolor{blue}{AM Weyl semimetal with MSC.}}-- 
The spin-resolved band structure of the monolayer Fe$_2$Te$_2$O without SOC is plotted in Fig. \ref{fig2}(c), from which  two remarkable  features can be observed. (i) The system exhibits an AM semi-metallic behavior, with both spin-up and spin-down channels featuring two WPs at the Fermi level. These WPs are located at the high-symmetry lines $X$-$M$ and $Y$-$M$ in the BZ, as illustrated in  Fig. \ref{fig3}(b).
(ii) The spin-up and spin-down states near the Fermi level are well separated from other bands and share the same eigenvalue of ${\cal{M}}_z$: $m_z=i$ for spin-up states and $m_z=-i$ for spin-down states, confirming the  monolayer Fe$_2$Te$_2$O is an ideal MSC material.

In the spin-up channel [Fig. \ref{fig2}(c)], the two 2D WPs  at the high-symmetry line $M$-$X$ are connected by  ${\cal{C}}_{2z}=\{C_{2z}^s||C_{2z}^l\}$, and are protected by $\left\{ E^s\Vert C_{2x}^l\right\}$, as the two bands forming the WPs respectively belong to band representations  $Y_2$ and $Y_3$, which have opposite eigenvalues of $\left\{ E^s\Vert C_{2x}^l\right\}$.
Since the  two WPs reside around $X$ point, we establish  a symmetry-allowed $k\cdot p$ Hamiltonian expanded from the $X$ point and based on the band representations $X_{1}^{-}$  and $X_{2}^{-}$, expressed as~\cite{refsm}
\begin{eqnarray}
	H_{X} & = & (m+t_xk_x^2+t_yk_y^2)\sigma_z+rk_xk_y\sigma_x,
\end{eqnarray}
where $\sigma$ denotes the Pauli matrices, all parameters are real, and $k_{x(y)}$ is measured from the $X$ point. According to the band structure, we have  $m=0.23 $ eV, $t_x = -0.025 \, \text{eV}\AA{^2}$, $t_y = -0.55 \, \text{eV}\AA{^2}$, and $r=0.42 \, \text{eV}\AA{^2}$.
When ${\bm k}=(0,\pm \sqrt{|m/t_y|},)$, the two bands cross linearly, forming two WPs $W_{\pm}^{\uparrow}$ [see Fig. \ref{fig3}(b)].
The effective Hamiltonian for the WP of  $W_{\pm}^{\uparrow}$ is obtained as 
\begin{eqnarray}\label{WPham}
	H_{\pm}^{\uparrow}& =&\pm\sqrt{|m/t_y|}(2t_yk_y\sigma_z+ r  k_x \sigma_x),
\end{eqnarray}
which indeed represents a  WP with linear dispersion along both $k_x$ and $k_y$.
The two spin-down WPs ($W_{\pm}^{\downarrow}$) are connected to $W_{\pm}^{\uparrow}$ by $\{E^s||C_{4z}^l\} \mathcal{T}$, and are protected by $\left\{ E^s\Vert C_{2y}^l\right\}$.

\textit{\textcolor{blue}{MCI  with MSC and quantized spin-Hall conductivity}}--
In the presence of SOC, monolayer Fe$_2$Te$_2$O belongs to magnetic space group No. 123.4.1002, which preserves the symmetries ${\cal{M}}_z$, $C_{4z}\mathcal{T}$ and spatial inversion $\mathcal{P}$, but breaks $\left\{ E^s\Vert C_{2x}^l\right\}$ and $\left\{ E^s\Vert C_{2y}^l\right\}$.
Thus, the four 2D WPs will be gapped by SOC.
Since each  gapped 2D WP gives  a half-integer Chern number  for the valence band~\cite{yao2009edge, liu2016pure}, the gapped Fe$_2$Te$_2$O  may become  an insulator with nontrivial band  topology.

The electronic band structure of the monolayer Fe$_2$Te$_2$O with SOC is shown in Fig. \ref{fig2}(d), where all the WPs  are gapped, consistent with the symmetry  analysis.
However, the other  remarkable feature (\textit{i.e.} MSC) of Fig. \ref{fig2}(c) persist [see Fig. \ref{fig2}(d)]. 
Specifically, while SOC induces a sizable band gap of $\sim$0.17 eV, it generates negligible spin hybridization, and the low-energy electrons with $m_z=\pm i$ still exhibit opposite and (almost) perfect persistent spin polarization.
Consequently, we can discuss the topology of the gapped WPs in different spin (mirror) subsystems. 

Under SOC effect, the Hamiltonian (\ref{WPham}) becomes~\cite{refsm}
\begin{eqnarray}\label{WPham2}
	{\cal H}_{\pm}^{\uparrow}& =&\pm\sqrt{|m/t_y|}(2t_yk_y\sigma_z+ r  k_x \sigma_x)+\Delta \sigma_y,
\end{eqnarray}
where $\Delta \sigma_y$ is the mass term caused by SOC effect and $\Delta\simeq0.103$ eV.
Particularly, the Chern number of the  gapped WP is $\frac{1}{2}\textrm{Sign}(t_yr\Delta)$, which is the same for both ${\cal{H}}_{+}^{\uparrow}$ and ${\cal{H}}_{-}^{\uparrow}$.
This  indicates that the SOC drives the spin-up ($m_z=i$) channel to  evolve into  a Chern insulator with ${\cal{C}}_+=1$.
Since the two spin channels are  connected by $C_{4z}\mathcal{T}$, the spin-down  ($m_z=-i$) channel should  also be a Chern insulator  but with  ${\cal{C}}_-=-1$.
This is further confirmed by the Wilson loop calculations performed on the two mirror subsystems, as shown in Fig. \ref{fig3}(a).
Therefore, the   monolayer Fe$_2$Te$_2$O  with SOC is an MCI with a mirror Chern number of ${\cal{C}}_{m}=({\cal{C}}_{+}-{\cal{C}}_{-})/2=1$.
To the best of our knowledge, monolayer Fe$_2$Te$_2$O is the first material candidate for a magnetic MCI.

A hallmark of MCI is the presence of topological gapless edge states, which have opposite mirror eigenvalues.
In Fig. \ref{fig3}(c), we plot the spectrum of a nanoribbon along $(1\bar{1}0)$ direction constructed by the monolayer Fe$_2$Te$_2$O  with SOC. 
Two counter-propagating edge states appear in the bulk gap, which intersects linearly at the $\bar{\Gamma}$ point.
Interestingly, these two edge states are almost entirely  contributed by the spin-up and spin-down electrons [see Fig. \ref{fig3}(c)].
While the system has SOC, the hybridization of the spin-up and spin-down of the edge states is always negligible, even in the region where the two states are close.
This is caused by the MSC. Without MSC,  the spin polarization of the  edge states is not persistent, but varies with the momentum.

Using a  four-terminal measurement setup, we study  the  spin transport of the  monolayer Fe$_2$Te$_2$O. 
The result is  shown in Fig. \ref{fig3}(d). One observes that the spin-Hall conductivity of the system is quantized to the well-known constant $e/(4\pi)$ in the bulk gap.
Therefore, we demonstrate that via the MSC, a quantized spin-Hall conductivity is  achieved  in an MCI rather than a QSH insulator.

\textit{\textcolor{blue}{Discussion.}}--
In this work, we have proposed a new coupling effect to realize quantized spin-Hall conductivity.
The underneath physics   shifts the search for QSH insulators to magnetic MCIs with MSC.

We highlight the robustness of the topological edge states in  Fig. \ref{fig3}(c).
The $\Gamma$-$M$ path in the bulk BZ corresponds to an effective 1D system, which have three mirrors: ${\cal{M}}_z$, ${\cal{M}}_{110}$, and  ${\cal{M}}_{1\bar{1}0}$. Due to  ${\cal{M}}_{1\bar{1}0}$, the 1D system can be classified by a $Z_2$ topological index, \textit{i.e.} the Zak phase. Moreover, the Zak phase can be further respectively calculated in the two ${\cal{M}}_{110}$-invariant subsystems. 
Interestingly, a detailed calculation shows  that the Zak phases of the $\Gamma$-$M$ path in the two ${\cal{M}}_{110}$-invariant subsystems are nontrivial, leading to two degenerate topological boundary states at the boundary, which exactly the $\bar{\Gamma}$ point in Fig. \ref{fig3}(c).
Thus, the gapless point at $\bar{\Gamma}$ is doubly  protected by ${\cal{M}}_z$ and ${\cal{M}}_{110}$, and then should be robust against the ${\cal{M}}_z$-breaking perturbations.

Furthermore, the band gap of monolayer Fe$_2$Te$_2$O with SOC is about 0.17 eV, which is much larger than the thermal energy at room temperature ($\sim 26$ meV). This suggests that the quantized spin-Hall conductivity can be observed at room-temperature, making the system promising for the design of room temperature devices.

Finally, we would like to point out that the search for realistic (SOC) materials with persistent spin polarization is  an important and active topic in the field of spintronics~\cite{schliemann2017colloquium, tao2018persistent, zhao2020purely, garcia2020canted, krol2021realizing, tian2017observation}. The MSC provides an alternative mechanism to realize it. 

\bibliography{ref}



\end{document}